%% file: main.tex
\documentclass[
superscriptaddress,
preprintnumbers,
amsmath,
amssymb,
prb,
twocolumn,
]{revtex4-2}

\usepackage[left=1.8cm, right=1.8cm, top=1.8cm, bottom=2.0cm]{geometry}
\usepackage[english]{babel}
\usepackage[utf8]{inputenc}
\usepackage{amsthm}
\usepackage{mathtools}
\usepackage{xcolor}
\usepackage[T1]{fontenc}
\usepackage{lipsum}
\usepackage{csquotes}
\usepackage{setspace}
\usepackage{float}
\usepackage{titlesec}
\titlespacing\subsubsection{0pt}{12pt plus 4pt minus 2pt}{0pt plus 2pt minus 2pt}
\setcitestyle{maxnames=5} %

\usepackage{tikz}
\usepackage{hyperref}

\usepackage{pgfplots}
\pgfplotsset{compat=1.18}
\usetikzlibrary{shapes,arrows,positioning,calc,decorations.pathmorphing,patterns,backgrounds,fit,3d,arrows.meta}

\begin{document}
\include{alias}

\title[Symmetry-restricted energy landscapes as a benchmark for machine learned interatomic potentials]{Symmetry-restricted energy landscapes as a benchmark for\\machine learned interatomic potentials}
\author{Abhijith S. Parackal}
%\email{abhijith.s.parackal@liu.se}

\affiliation{Department of Physics, Chemistry and Biology, Linköping University, Linköping, Sweden}
\author{Rickard Armiento}
\email{rickard.armiento@liu.se}
\affiliation{Department of Physics, Chemistry and Biology, Linköping University, Linköping, Sweden}

\author{Florian Trybel}
\email{florian.trybel@liu.se; RA and FT contributed equally to this work}
\affiliation{Department of Physics, Chemistry and Biology, Linköping University, Linköping, Sweden}

\begin{abstract}
Machine learned interatomic potentials (MLIPs) are becoming a standard method for DFT-level accurate molecular dynamics simulation and large-scale studies of crystal energetics. Increasingly popular are universal pre-trained potentials, also called foundation models, based on, e.g. the MACE, CHGNet, M3GNet, ORB, and SevenNet architectures.
While there are many benchmarks of these models using validation errors and materials discovery tasks, their fidelity in reproducing the detailed features of potential energy surfaces (PES) is not understood to the same degree.
We evaluate the accuracy of these potentials by systematically probing their predicted energy landscapes.
Two-dimensional slices of the potential energy surface are constructed where the atomic positions are varied along selected Wyckoff degrees of freedom within a fixed crystal symmetry.
This approach enables a direct, visual comparison of the interatomic potentials and DFT-calculated surfaces which reveals potential artifacts, \textit{e.g.}, arising from unique local environments.
Our analysis highlights the strengths and limitations of different potentials in capturing local minima, saddle points, and overall PES topology, offering insights into the physical accuracy of current pre-trained IAPs and providing benchmarks for future model development.
\end{abstract}

\maketitle

\section{Introduction}

The potential energy hyper-surface (PES), also known as the potential energy landscape, represents a high-dimensional object that maps all degrees of freedom (Cartesian coordinates) of a crystal structure to an energy  \cite{smeetonVisualizingEnergyLandscapes2014a}, conventionally assuming a static picture (no temperature or zero-point motion) at a fixed composition. Minima in this landscape represent structural configurations with an energy barrier to other configurations. The height of these barriers gives an indication of the possible \mbox{(meta-)}stability of a configuration and can be used to rationalize reaction pathways using, \textit{e.g.}, nudged elastic band calculations. Accurate modeling of the PES is essential for understanding and predicting a variety of material behaviors at the atomic level, such as elastic properties, diffusion barriers, and lattice vibrations~\cite{walesEnergyLandscapes2001}. 

Until recently, density functional theory (DFT) \cite{Hohenberg1964,Kohn1965} has been the main tool to reliably model the PES, as it provides a detailed quantum mechanical description of the electronic structure. 
However, especially for larger systems, the computational cost of DFT becomes significant, making this a prohibitively expensive way to map the relevant regions of the PES~\cite{Heinen_2020,yuSystematicAssessmentVarious2024a}. Hence, it is attractive to turn to interaction potentials. Classical interatomic potentials are constructed by fitting either just simple pairwise analytical functions, such as Lennard-Jones or Morse expressions, or a few higher-order many-body terms (\textit{e.g.}, EAM, Tersoff, Stillinger–Weber) to reproduce energies, forces, or structural properties obtained from first-principles calculations or experiments, see Ref.~\cite{Müser31122023} for a review. Such potentials have been successfully used to accelerate energy calculation as they scale well up to large simulation cells, but their generation is complex, and the potentials are in general not transferable beyond a specific application. 

Machine-learned interatomic potentials (MLIPs) have been shown to overcome the limitations of traditional methods, with the ability to achieve DFT-level accuracy with significantly improved computational efficiency for some applications.
Universal pre-trained MLIPs (\umlips) based on, \textit{e.g}., the MACE~\cite{mace}, CHGNet~\cite{chgnet}, and ORB~\cite{orb_v2} architectures are trained on DFT relaxation trajectories extracted from large datasets (most commonly from Materials Project \cite{jainCommentaryMaterialsProject2013, chgnet}, and more recently, Alexandria~\cite{schmidtMachineLearningAssistedDeterminationGlobal2023}), enabling them to approximate the PES with high fidelity for a wide range of chemical compositions.
The capabilities of these types of potentials to sufficiently accurately describe energetics have greatly expanded the capabilities of computational materials science, making it feasible to investigate complex material systems that were previously too resource-intensive~\cite{zuoPerformanceCostAssessment2020}.

While \umlips\ achieve low validation errors across a wide range of available benchmarks~\cite{riebesellMatbenchDiscoveryFramework2024} and are successfully applied in various applications, including the large scale discovery of novel materials \cite{merchantScalingDeepLearning2023, yangMatterSimDeepLearning2024} and large cell-size thermodynamic modeling \cite{erhardModellingAtomicNanoscale2024}, there remains uncertainty regarding their ability to accurately capture the underlying physics of the material systems and local environments not part of the training data~\cite{Wen2020}.
The description of regions in the energy landscape far from the global minimum is an inherently difficult endeavor for interaction potentials \cite{Tran2017}, and the extent to which the pre-trained potentials can sufficiently replicate intricate PES features, such as high-energy local minima, saddle points, and gradients far from equilibrium, is not fully understood. Recent works have highlighted challenges in the generalization capabilities of these models \cite{huangCrossfunctionalTransferabilityUniversal2025, kaplanFoundationalPotentialEnergy2025}.
For instance, Deng \textit{et al.} \cite{deng2024overcomingsystematicsofteninguniversal} observed a consistent softening of the potential energy surfaces far from minima.
Discrepancies are often attributed to biased sampling of near-equilibrium atomic arrangements in pre-training datasets, leading to systematic under-prediction errors in the local PES curvature around minima. 
%These findings suggest that, despite their computational efficiency and success in a wide range of applications, \umlips\ may struggle to accurately capture complex PES features, particularly in scenarios involving configurations far from equilibrium.
Several studies that use \umlips\ for crystal structure discovery report instances where a converged geometry optimization performed with \umlips\ leads to unphysical structures, where DFT-based optimization fails to even converge~\cite{merchantScalingDeepLearning2023,liuDiscrepanciesErrorEvaluation2023}.
The present authors have made similar observations when applying uMLIPs for materials discovery \cite{parackalScreening39Billion2026}, motivating a deeper investigation into the limitations of \umlips\ in faithfully representing the PES topology. 

Understanding and addressing the limitations of uMLIPs is crucial for the reliable application of these techniques for modeling material behaviors that are sensitive to the precise details of the PES. 
A common method is to use fine-tuning~\cite{kaurDataefficientFinetuningFoundational2025} or transfer-learning~\cite{radovaFinetuningFoundationModels2025} of pre-trained \umlips\ for targeted applications.
These strategies provide a highly data-efficient way to obtain high-accuracy potentials for applications outside the initial pre-training data coverage.
However, fine-tuning may have unintended side effects \cite{kumarFineTuningCanDistort2022}, motivating a careful investigation of how the ability to describe the PES changes as a result of applying these techniques.

Visualization provides a way to understand the features and inaccuracies of potentials that can be far more informative than an opaque value meant to digest the accuracy across a wide range of applications, or across the phase space. Since direct visualization of high-dimensional PES is not feasible, a variety of dimensionality reduction methods~\cite{PhysRevX.11.041026, bihani2024lowdimensional}, including connectivity maps to connect local minima~\cite{smeetonVisualizingEnergyLandscapes2014a, walesDecodingEnergyLandscape2012}, are used. For example, Yunsheng Liu \textit{et al.} recently studied the behavior of a number of MLIPs for the energy landscape of one of the vibrational modes in Silicon, revealing discrepancies in the energy landscape obtained with DFT~\cite{liuDiscrepanciesErrorEvaluation2023}. %Similarly, Lenz \textit{et al.} demonstrated  that the PES of cubic \chem{ZrO_2} in a reduced parameter space {\color{red}[Some word is missing here?]} \cite{lenzParametricallyConstrainedGeometry2019}. 
However, visualizations like this typically need to be tailored to represent the features of a specific system and highlight the relevant information, which 
can make them difficult to interpret without detailed understanding of the analysis method. 
%the large amount of information in visualizations of the PES can make them difficult 
\begin{table*}[t]
\begin{center}
    \caption{Details of the pretrained interatomic potential models used in this study.}
    \begin{tabular}{l @{\hspace{0.5cm}} l @{\hspace{0.5cm}} c}
        \hline
        \textbf{Model} & \textbf{Weights}  & \textbf{Source code version} \\
        \hline
        % Example entries:
        \macemedium & \texttt{MACE-MP-0 small} \cite{mace, mace_mp_0} & 0.3.8 \\
        \maceoba &  \texttt{mace\_mp\_0b2} \cite{mace,mace_mp_0b2}& 0.3.8 \\
        \maceobb &  \texttt{mace\_mp\_0b3} \cite{mace,mace_mp_0b3}& 0.3.8 \\
        \macempa &  \texttt{mace\_mpa\_0} \cite{mace,mace_mpa_0}& 0.3.8 \\
        \maceomat &  \texttt{mace\_omat\_0} \cite{mace,mace_omat_0,OMat24}& 0.3.8 \\
        \macematpes &  \texttt{mace\_matpes\_0} \cite{mace,mace_matpes_0,kaplanFoundationalPotentialEnergy2025}& 0.3.8 \\
        \seventa & \texttt{SEVENNET\_0\_11Jul2024} \cite{sevenet, sevenet_pretrained_0} & 0.10.1 \\
        \orb & \texttt{orb-v2} \cite{neumannOrbFastScalable2024, orb_v2} & 0.4.0 \\
        \chgnet & \texttt{model 0.3.0} \cite{chgnet,chgnet_030} & 0.3.8\\

        % Add more rows as needed
        \hline
    \end{tabular}
    \label{tab:model_summary}
    \end{center}
\end{table*}

In this paper, we demonstrate a way to apply visual analysis of slices of the PES to (i) explore how the ability of MLIPs to reproduce specific features is connected to their description of different types of physics, and (ii) compare and refine MLIPs. We present a complete workflow to yield symmetry constrained two-dimensional PES (s2DPES) as a function of the degrees of freedom allowed by the symmetry of a crystal structure. This is achieved by representing the crystal structure via its Wyckoff positions \cite{ITA2002}, which describe the symmetry-related atomic sites within the crystal, providing a systematic method to explore atomic configurations and their corresponding potential energy surfaces along symmetry-allowed changes of the positions of the atoms.
The pairwise mapping of Wyckoff degrees of freedom into a 2D energy landscape enables a very direct comparison between the PES predicted by an \umlips\ and the corresponding one from DFT calculations.
This allows us to assess the accuracy of each model in capturing even the smallest critical features of the PES, such as local minima, saddle points and any artifacts that may arise from the unique local environments of the structures. The s2DPESs also provide direct access to the topology of the landscape, which may be useful for other techniques in visual and automated analysis.
%could be useful for symmetry breaking under phase transition, active learning" or something .

% \section{Background}
% \subsection{Symmetry constrained energy landscapes} 
%motivation,why symmetry constrained energy landscapes are useful

\section{Background}

\subsection{Symmetry-allowed degrees of freedom}

For a given space group, Wyckoff positions define sets of symmetry-equivalent atomic sites that are related through the symmetry operations of the space group \cite{wyckoff1922analytical}.
As a consequence, a single Wyckoff position encodes the positions of multiple atoms \cite{goodallRapidDiscoveryStable2022b}, referred to as the multiplicity of the Wyckoff position.
This implies that any modifications to the degrees of freedom associated with a Wyckoff position (whether constrained to a point, line, plane, or fully unconstrained in three-dimensional space) simultaneously affect all symmetry-equivalent atoms.

The evaluation of the effects of the local symmetry of a structure is useful for studying defect-, piezoelectric-, and other electronic structure-related properties \cite{lenz2019parametrically}.
Taking advantage of crystal symmetry significantly reduces the dimensionality of the configuration space.
Instead of the initial $6+3N$ parameters for a system with $N$ atoms, only the lattice parameters and the Wyckoff degrees of freedom need to be considered.
This approach is useful for identifying the ground-state crystal structure in a given symmetry more efficiently by relaxing the structure only along the free parameters of symmetry \cite{Wang2018, Reinaudi2000, lenz2019parametrically}.

To reference specific crystal structure systems without specifying the Wyckoff degrees of freedom we use \textit{protostructure labels} \cite{xrd_paper}, which are based on AFLOW prototype labels \cite{AFLOWLibraryCrystallographic}. They consist of a combination of the anonymized chemical formula, the Pearson symbol, the space group, the occupied Wyckoff positions, and the element occupying the respective Wyckoff position.
For example, \texttt{AB3C\_tP40\_135\_f\_3g\_h:Al-N-Ti} describes a structure with the chemical formula $\mathrm{AlTiN}_3$, Pearson symbol tP40, and where Al atoms occupies the Wyckoff position labeled \texttt{f}, N atoms occupies 3 different orbits represented by Wyckoff position \texttt{g}, and Ti occupies Wyckoff position labeled \texttt{h} in space group 135.

\subsection{Pretrained interatomic potentials}
\label{sec: pretrained interatomic potentials}
The MLIPs of primary interest in this work %Machine learned interatomic potentials are a class of empirical potentials that 
are trained on \textit{ab-initio} electronic structure data (typically 
energies, forces, and stress).
One class of such potentials uses a graph neural network to describe atomic interactions, with some architectures explicitly incorporating invariance/equivariance features and graph convolutions, leading to extremely high data efficiency as well as physically informed models.
Such models can be trained on existing DFT data from large databases (\textit{e.g.}, Alexandria \cite{doi:10.1126/sciadv.abi7948,schmidt2024,schmidtMachineLearningAssistedDeterminationGlobal2023}, Materials Project \cite{jainCommentaryMaterialsProject2013}) to accurately reproduce equilibrium structures across a wide range of compositions.
The success of this approach paved the way for pre-trained interatomic potentials, where the weights from training the potential on a broad data set are distributed along with the potential, with the aim that they can be used for general structures without additional training. In benchmarks, recent such uMLIPs give near density functional theory level accuracy for geometry optimization at orders of magnitude higher computational efficiency~\cite{riebesellMatbenchDiscoveryFramework2024}.
The field of interatomic potentials is very active, with new models continuously appearing. The examples shown in this work are only quantitatively applicable for the reported model weights, while the suggested workflow can be applied universally. 

In this work, we predict energies of s2DPES for a range of uMLIPs. In particular, we explore ORB, CHGnet, Sevennet, and several versions of MACE models. The MACE models are a class of equivarient message passing graph neural networks that use the atomic cluster expansion to capture 4-body interactions.
The message passing architecture allows the model to essentially capture ~10+ body order \cite{macebodyorder}.
Since its conception, the MACE architecture has been periodically updated and improved, and the models available are trained on successively increasing amounts of data.
The model referred to in this work as \macemedium\ originates from the first iteration of MACE models~\cite{mace} trained on the relaxation data from the Materials Project database (\textit{i.e.}, the MPTraj \cite{chgnet, jainCommentaryMaterialsProject2013} dataset).
Subsequent models incorporate, \textit{e.g.}\ a repulsion term at small inter-atomic distances (\maceoba\ and \maceobb). We use the following label for the different models:
\macempa\ is trained on the larger Alexandria dataset~\cite{schmidtMachineLearningAssistedDeterminationGlobal2023}, while \maceomat\ is trained on the OMat24~\cite{OMat24} dataset.
\macematpes\ is \maceomat\ model but fine tuned on the MATPES~\cite{kaplanFoundationalPotentialEnergy2025} dataset.

The ORB architecture is based on a smoothed graph attention mechanism~\cite{neumannOrbFastScalable2024}. 
Specifically, the ORB models use a unique training strategy, starting with a de-noising diffusion model that was trained on crystal structure data. The base model was subsequently trained to produce the energies, forces, and stress in a supervised learning fashion from the MPTraj~\cite{chgnet} dataset.
Also, unlike other models presented here, forces and stresses are directly predicted rather than calculated using the autograd capability of modern machine learning architectures, which makes the architecture non-conservative, but significantly faster at inference~\cite{neumannOrbFastScalable2024}.

\input{introduction_figure}

The potential labeled \chgnet\ is a graph neural network-based architecture that additionally captures the on-site magnetic moments to enable charge-informed atomistic modelling \cite{chgnet}.
SevenNet \cite{sevenet} is a Nequip-based \cite{batznerE3equivariantGraphNeural2022} graph neural network that uses efficient spatial decomposition to achieve ideal strong-scaling performance. We used the first iteration of the model named \seventa.
All calculations were performed on GPUs with float32 precision.
The pre-trained models employed in this study are summarised in Table~\ref{tab:model_summary}.
% \todo{change the table so its model, weights, data, source code version}.

\section{Methods}

\subsection{Distance function}

When atoms are positioned too closely within the unit cell, they create an unphysical environment that most interatomic potentials have not been trained to handle.
Since such structures generally have very high DFT energies, they are rarely represented in the training data sets. As a consequence, the potentials may yield significant errors for systems with atoms in very close proximity \cite{donaldson2024genetic}. Some interatomic potentials address this issue by incorporating an explicit
repulsion term \cite{Eyert2023, mace}.

In our workflow, we track such atomic configurations by defining a minimum distance threshold.
Specifically, we use the sum of the Wigner-Seitz radii of two atoms, denoted as $D_{\text{min}}$, as the minimum allowed distance between any two atomic pairs.
In general, when comparing s2DPES, only configurations where all pair distances exceed this threshold are considered meaningful.

A simple cost function for the minimum distance is defined as
\begin{equation}
    C_{\text{distance}} = \frac{1}{2}\sum_{i=1}^{n}\sum_{j=1}^{n}\max\left(0,\;D^{\min}_{(i,j)} - \left(D_{(i,j)} + d_{\text{lat}}^{\min}\delta_{ij}\right)\right),
\label{eqn:distance_eq}
\end{equation}
where $D_{i,j}$ is the smallest distance between atoms $i$ and $j$, computed including periodic boundary conditions.
To prevent self-distances of zero, the term $d_{\text{lat}}^{\min}$, which is the shortest distance between an atom and its periodic image (excluding the origin), is added when the two atoms are the same.

\subsection{Workflow}
\label{sec: workflow}

Our workflow for producing s2DPES is illustrated in Fig.~\ref{fig:graph_repr}. It begins with the user providing a crystal structure file. This input structure is processed with the Python library \symgen~\cite{xrd_paper, parackalScreening39Billion2026}, which provides a range of utilities related to crystal structures and symmetries,  interoperable with ASE (Atomic Simulation Environment)~\cite{ase-paper} to support a wide range of file formats. Created structures are standardized and symmetrized using \texttt{spglib}~\cite{spglibv2} to identify Wyckoff positions and their corresponding DOF.

\begin{figure*}[t!]
    \includegraphics[width = 0.95\textwidth]{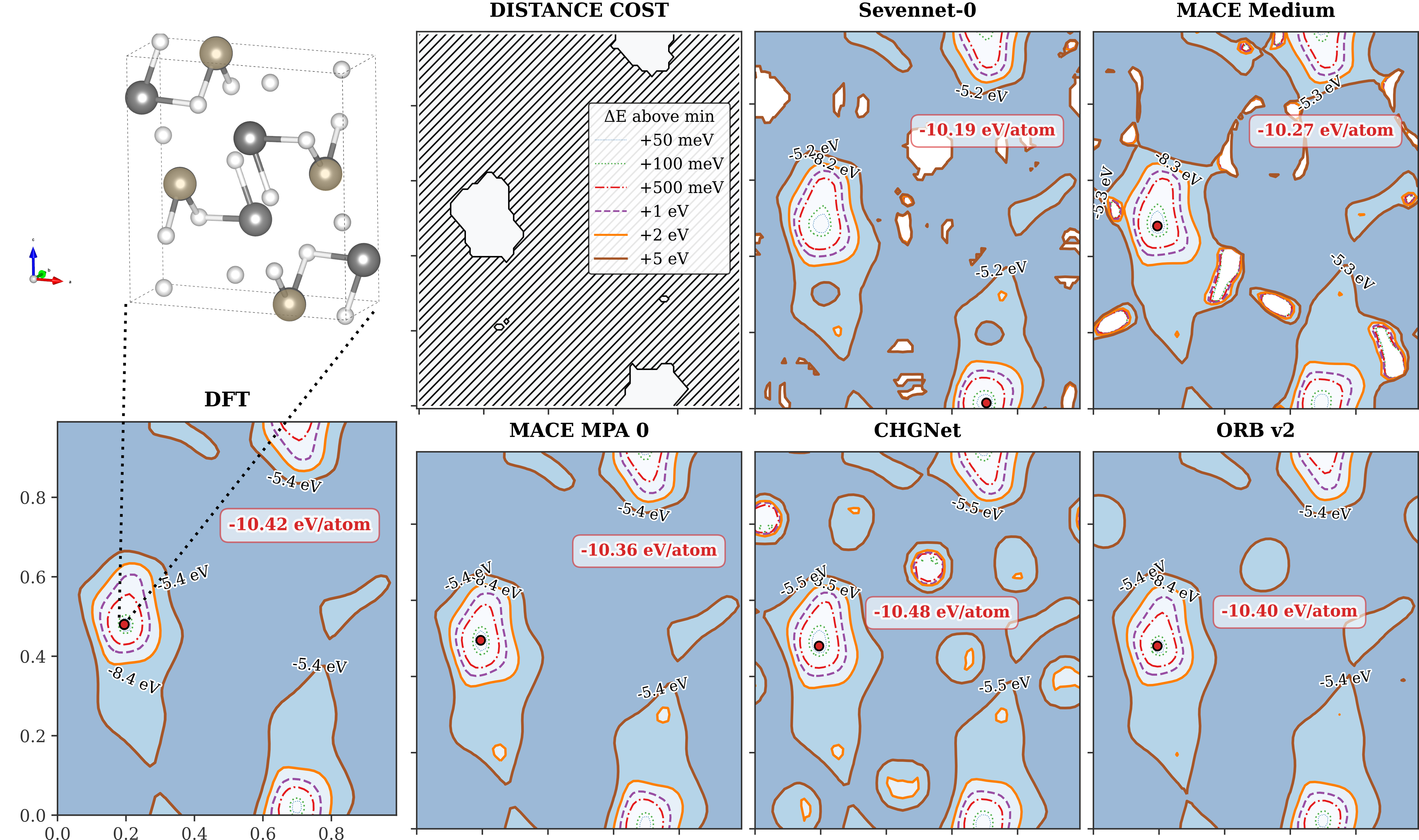}
    \caption{Two-dimensional energy landscape for $\mathrm{W}_2\mathrm{N}_3$ generated by varying the \textit{x} and \textit{z} Wyckoff degrees of freedom of tungsten atoms (indicated with blue colour in the crystal structure figure) in the protostructure \protoid{A3B2_oP20_62_3c_2c:N-W}. The distance cost is calculated with a minimum interatomic distance of 1.1115 \AA\ for N–N pairs, 1.647 \AA\ for N–W pairs, and 2.1825 \AA\ for W–W pairs. Energy values are presented in eV/atom.\label{fig:W2N3}}
\end{figure*}

Based on the symmetry analysis of the input structure, the \texttt{StructureSolver} class operates on the Wyckoff position of each representative site and generates fully filled cell representations from a list of values for the degrees of freedom (DOF). 
The user can specify which  DOFs for a given Wyckoff position should be sampled. This involves selecting specific Wyckoff positions and specifying the desired positional variations, such as permitting movement along \textit{e.g.}, the \textit{x} and \textit{z} directions while keeping the \textit{y} coordinate fixed, given the Wyckoff position has all 3 degrees of freedom. The library automatically performs sanity checks on Wyckoff positions to ensure that only actual DOF are sampled.  While the program allows varying an arbitrary number of DOFs, for the s2DPES we always vary two per structure.

In the workflow, the selected degrees of freedom are varied systematically to produce atomic configurations. The user may define the range and step size for these variations, which are used to generate a grid of atomic configurations.
 To maintain physical plausibility, the distance cost function, Eq.~\ref{eqn:distance_eq}, is used to penalize configurations where atoms come too close to each other. %This function calculates the penalty for any atomic pair that violates predefined minimum interatomic distance thresholds, as defined in Eq.~\ref{eqn:distance_eq}. 
 The full workflow is implemented using JAX~\cite{jax2018github} and allows for highly efficient vectorized computations on all modern GPU architectures.

For each generated structure, the potential energy is computed using various 
 uMLIPs, based on, \textit{e.g.}\ MACE, CHGNet, ORB, and SevenNet architectures. These energy calculations are performed in parallel batches to optimize computational efficiency. The resulting energies are used to construct two-dimensional energy landscapes, which are visualized using contour plots.
These plots highlight regions around local minima and saddle points, with the reference structure, initially provided by the user, marked for comparison. This systematic approach enables a detailed evaluation of the energy landscapes predicted by different models.

This workflow allows for a systematic and efficient exploration of 2D submanifolds of the PES by varying atomic positions within the constraints of the crystal symmetry.
By comparing the energy landscapes generated by, \textit{e.g.}, different models, the accuracy of these models in capturing intricate features of the PES, such as local minima and saddle points, can be assessed. The generated s2DPES plots are easy and fast to interpret visually by the user, and can be produced for a wide range of structures. A more detailed analysis of the local curvature allows to efficiently analyze difference in model behavior. In particular, the effects of fine-tuning or transfer learning, modifications to the training data, or even small changes to the model architecture can be efficiently tracked and easily communicated to non-experts.

\section{Results}
 In the following, examples of s2DPES plots are provided for different crystal structures and models, highlighting different possible analysis pathways enabled by the library. 
 
 \begin{figure*}
\begin{center}
    \includegraphics[width = 1.0\textwidth]{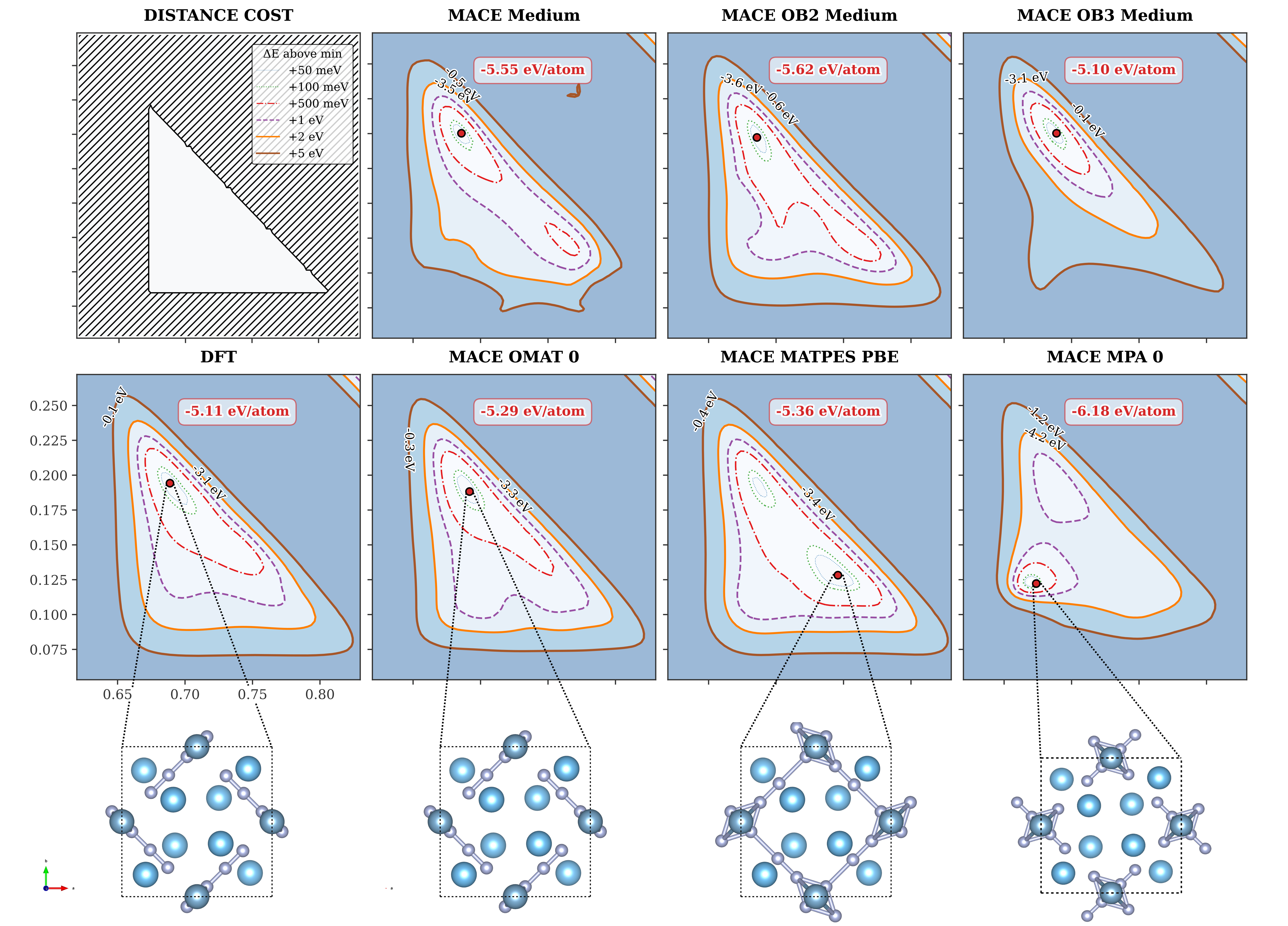}
    \caption{Energy landscapes for $\mathrm{Al}\mathrm{Ti}\mathrm{N}_{3}$ from $x$-direction mutations of two N atoms in \protoid{AB3C_tP40_135_f_3g_h:Al-N-Ti},  which was examined after inconsistent outputs between \macempa\ and other MACE models.
Minimum interatomic distances used for the distance cost are 1.98 \AA\ (Ti–Ti), 1.55 \AA\ (Ti–N), 2.04 \AA\ (Ti–Al), 1.112 \AA\ (N–N), 1.61 \AA\ (N–Al), and 2.10 \AA\ (Al–Al).
Energies are in eV/atom.\label{fig:tialn}}
\end{center}
\end{figure*}

\subsection{W\textsubscript{2}N\textsubscript{3}}
\label{sec: W2N3}

Meta-stable phases recovered from high-pressure high-temperature synthesis are
particularly challenging to model. These materials often show unexpected chemistry and complex bonding. Recently, two tungsten nitrides phases with a wide range of technologically relevant properties (\textit{e.g.}, ultra-incompressible, superhard, and superconducting) have been synthesized in diamond anvil cells and recovered to ambient conditions~\cite{Liang2024}. The synthesized \chem{W_2N_3} is so far not represented in any training corpus of the uMLIPs considered here, which makes it an excellent test case for evaluating the predictive performance of machine-learned interatomic potentials.

The crystal structure belongs to space group \textit{Pnma}, with Nitrogen and Tungsten atoms occupying Wyckoff positions \texttt{3c} and \texttt{2c}, respectively.
The protostructure label is \protoid{A3B2_oP20_62_3c_2c:N-W}.
The Wyckoff position \texttt{c} allows two degrees of freedom, along the \textit{x} and \textit{z} directions.
As shown in Figure~\ref{fig:W2N3}, we choose to relax the Wyckoff positions of the tungsten atoms, allowing displacements along these two positional degrees of freedom.
This generates a two-dimensional energy landscape useful for evaluating model predictions.
Across most models, the resulting energy landscapes are smooth and consistent, accurately identifying the local energy minimum at the equilibrium position of the tungsten atom.
Additionally, all models successfully capture the general curvature of the potential energy surface near this minimum.

\begin{figure*}[t!]
    \includegraphics[width=0.95\textwidth]{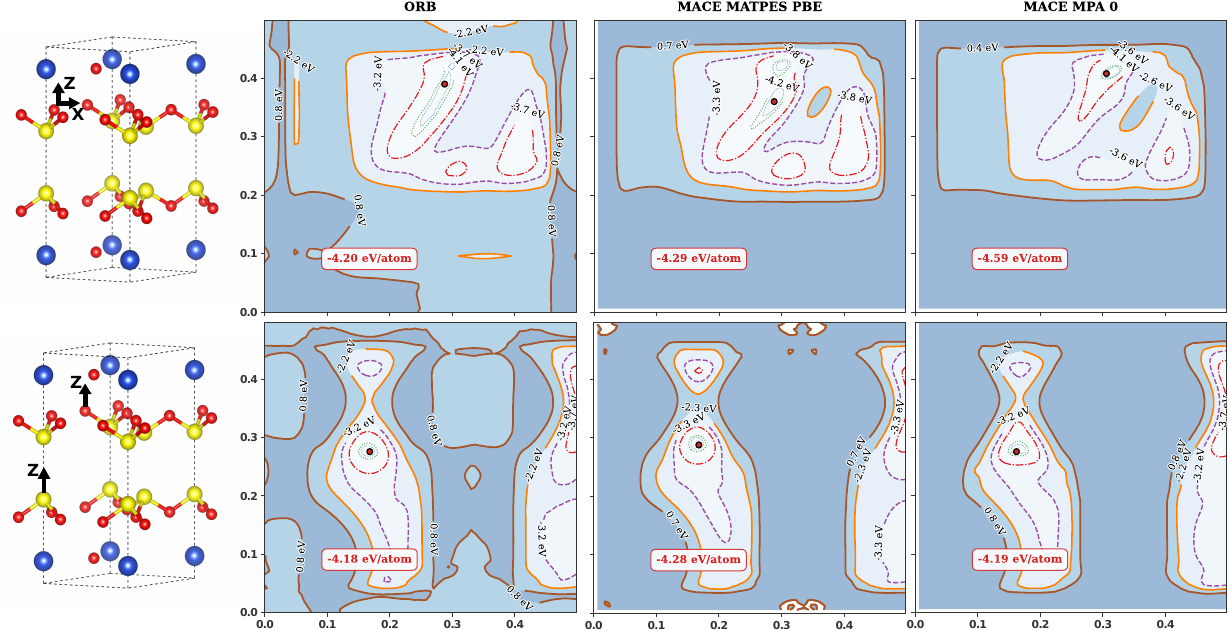}
    \caption{
Energy landscapes of \chem{Cu_2O_8S_4} with protostructure label \protoid{A4B2C_hP14_187_g_in_gh:Cu-O-S} along two different degrees of freedom.
     (a) Two degrees of freedom  along \textit{x} and \textit{z} of Oxygen at Wyckoff position \textit{n},
    (b) Oxygen in Wyckoff position \textit{n} and Sulfur in Wyckoff position \textit{g} perturbed along \textit{z}.}
\label{fig:Cu2O8S4}
\end{figure*}

However, some notable issues emerge.
\seventa\ and \chgnet\ display artifacts or noisy behavior in regions where atomic overlap becomes significant, characterized by an unphysical drop in predicted energy.
This behavior is commonly observed in models that lack an auxiliary repulsion term, as such configurations are typically absent from training datasets. \macemedium\ also shows additional artifacts, even in regions where the minimum interatomic distance is not violated.
These artifacts could potentially disrupt optimization processes if the initialization occurs near these regions.

Interestingly, \orb\ also predicts an energy drop in regions of close atomic proximity, although less dramatically than the other models.
Furthermore, \orb\ predicts a notably narrower energy range, with a difference of approximately 8 eV/atom between the highest and lowest energies, significantly smaller than ranges observed in other models.

A main conclusion from the s2DPES is that all tested models accurately capture the local energy minimum for a structure outside the training data.

\subsection{AlTiN\textsubscript{3}}

One of the outcomes of the crystal structure search in Ref.~\cite{parackalScreening39Billion2026} was a structure for AlTiN\textsubscript{3} with protostructure label \protoid{AB3C_tP40_135_f_3g_h:Al-N-Ti}. The subsequent analysis revealed discrepancies in the predicted energy landscapes between \macempa\ and other MACE models, leading to the structure being examined more closely for validation.
The structure possesses six Wyckoff degrees of freedom.
Here, we focus on the configurational space defined by mutations of two nitrogen groups occupying the Wyckoff site 'g', specifically varying their positions along the $x$ direction to generate a two-dimensional energy landscape.

Figure~\ref{fig:tialn} presents a detailed view of the energy landscape in the vicinity of the initially reported structure. The comparison reveals pronounced discrepancies between the potential energy surfaces predicted by \macempa\ and those obtained from other MACE models and from reference DFT calculations. Most notably, \macempa\ predicts a distinct local minimum in a region where both DFT and other MACE models indicate no stable configurations.
With the aid of the s2DPES, it is thus clear that the geometry optimization using the \macempa\ model and the ASE \cite{ase-paper} gets trapped in the spurious basin that only appears in this version of MACE. In the region near the minimum that is found when optimizing the structure with the other potentials, however, it is notable that \macempa\ reproduces the landscape reasonably well, suggesting that the erroneous region is a model-specific artifact that appears only away from the global minimum. However, as seen here, such features can distort the outcome of a relaxation. 

Our result for this system highlights the benefit of using models trained on different data sets and preferably also based on different architectures for, \textit{e.g.}, applications in structure search.
Most of the MACE models demonstrate robust and reliable performance, accurately capturing the topology of the potential energy surface.
Furthermore, we observe a clear progression in model quality across generations: the newer \maceomat\ model yields an energy landscape that closely matches the DFT reference, demonstrating the benefits of improved training data and architectural refinements. 

\subsection{Cu\textsubscript{2}O\textsubscript{8}S\textsubscript{4}}

This structure was selected to demonstrate a notable discrepancy between \macematpes\ and other models during symmetry-constrained relaxation. Starting from a randomly initialized configuration within the allowed Wyckoff degrees of freedom, all models except \macematpes\ relaxed into a common structure, while \macematpes\ converged to a distinct local minimum. 

To investigate this behavior, we used the s2DPES workflow to investigate two-dimensional energy landscapes over three relevant degrees of freedom: (a) the \textit{x} and \textit{z} coordinates of Oxygen at Wyckoff position \textit{n}, and (b) the \textit{z} coordinates of Oxygen at \textit{n} and Sulfur at \textit{g}. As shown in Fig.~\ref{fig:Cu2O8S4}, \macematpes\ consistently identifies a different local minimum compared to the other models across both cases.
The alternative minimum predicted by \macematpes\ persists even when symmetry constraints are lifted and a full relaxation is performed, indicating that this is not merely an artifact of the symmetry-constrained landscape.
The presence of multiple minima in the landscape may also have implications for structure prediction and phase stability assessments using machine-learned potentials.

\section{Conclusions}
We have presented a workflow for straightforward visualization of two-dimensional potential energy surfaces (PES) constrained by crystal symmetry. We have demonstrated how it can be used for systematic visual evaluation of state-of-the-art pretrained interatomic potential models via direct comparison between model predictions and DFT reference calculations, revealing both strengths and limitations of the models.
While most models accurately capture local minima and reproduce the general PES curvature near the equilibrium, significant discrepancies and artifacts can arise, particularly in regions far from equilibrium or in the presence of unphysical atomic arrangements. Notably, some models exhibit erroneous minima or fail to penalize small interatomic distances, underscoring the importance of explicit repulsion terms and comprehensive training data. 

The PESs of the examples used to demonstrate our workflow highlight the utility of symmetry-constrained landscape visualization for diagnosing model failures, understanding metastability, and identifying regions of uncertainty--insights that are valuable for active learning and model improvement. The observed differences between models and across generations further emphasize the need for rigorous benchmarking beyond simple scalar error measures. % standard validation errors.
This work provides a robust framework for assessing and improving the physical fidelity of interatomic potentials, with implications for reliable structure prediction, phase stability analysis, and the broader application of machine-learned potentials in materials discovery. With the rapid development towards increasingly sophisticated and complex architectures, we foresee a growing need for robust general tools to explore and diagnose features of interaction potentials of the kind presented here.

\section{Acknowledgment}
R.A. acknowledges financial support from the Swedish Research Council (VR) through Grant No. 2020-05402 and the Swedish e-Science Research Centre (SeRC). F.T. acknowledges support through ERC Grant (UNMASCC-HP, 101117758) and the Swedish government’s Strategic Research Area in Materials Science on
Functional Materials at Linköping University (faculty grant SFO-Mat-LiU 2009-00971). The computations were enabled by resources provided by the National Academic Infrastructure for Supercomputing in Sweden (NAISS), partially funded by the Swedish Research Council through grant agreement no. 2022-06725. We also acknowledge NAISS for providing access to
the LUMI supercomputer, owned by the EuroHPC Joint
Undertaking and hosted by CSC (Finland) and the LUMI
consortium.

\newpage
\bibliographystyle{apsrev4-2}
\bibliography{references.bib}
\end{document}

%% file: alias.tex
\definecolor{todoColor}{HTML}{E74C3C}
\definecolor{fixmeColor}{HTML}{D35400}
\definecolor{noteColor}{HTML}{2ECC71}
\definecolor{questionColor}{HTML}{8E44AD}
\definecolor{highlightColor}{HTML}{F1C40F}

\newcommand{\todo}[1]{\textcolor{todoColor}{\texttt{TODO:} #1}}
\newcommand{\fixme}[1]{\textcolor{fixmeColor}{\texttt{FIXME:} #1}}
\newcommand{\note}[1]{\textcolor{noteColor}{\texttt{NOTE:} #1}}
\newcommand{\question}[1]{\textcolor{questionColor}{\texttt{?} #1}}
\newcommand{\highlight}[1]{\colorbox{highlightColor}{#1}}

%% abhi
\newcommand{\protoid}[1]{\texttt{\detokenize{#1}}}
\newcommand{\chem}[1]{\ensuremath{\mathrm{#1}}}

\newcommand{\symgen}{\texttt{httk-symgen}}
\newcommand{\umlips}{uMLIPs}
\newcommand{\macemedium}{\texttt{MACE\_medium}}
\newcommand{\maceoba}{\texttt{MACE\_OB2}}
\newcommand{\maceobb}{\texttt{MACE\_OB3}}
\newcommand{\macempa}{\texttt{MACE\_MPA-0}}
\newcommand{\maceomat}{\texttt{MACE\_OMAT-0}}
\newcommand{\macematpes}{\texttt{MACE\_MATPES-PBE}}

\newcommand{\chgnet}{\texttt{CHGNet}}
\newcommand{\seventa}{\texttt{SevenNet\-0}}
\newcommand{\orb}{\texttt{ORB v2}}

%% file: introduction_figure.tex
\pgfplotsset{compat=1.18}
\definecolor{atomTypeA}{RGB}{65, 185, 65}    % Green
\definecolor{atomTypeB}{RGB}{200, 65, 65}    % Red
\definecolor{atomTypeC}{RGB}{80, 120, 220}   % Blue
\definecolor{potentialHot}{RGB}{255, 180, 0} % Yellow
\definecolor{potentialMid}{RGB}{230, 80, 80} % Red
\definecolor{potentialCold}{RGB}{30, 50, 120}% Dark blue

\begin{figure*}[t!]
\begin{center}

\begin{tikzpicture}[scale=0.75]

    %Part 1: 3D Crystal Structure with Wyckoff positions
    \begin{scope}[shift={(-0.0,0.5)}, scale=0.75]
        % Unit cell
        \draw[black, thick] (0,0,0) -- (3,0,0) -- (3,3,0) -- (0,3,0) -- cycle;
        \draw[black, thick] (0,0,0) -- (0,0,3) -- (3,0,3) -- (3,0,0);
        \draw[black, thick] (0,0,3) -- (0,3,3) -- (3,3,3) -- (3,0,3);
        \draw[black, thick] (0,3,0) -- (0,3,3);
        \draw[black, thick] (3,3,0) -- (3,3,3);
        
        % Atoms Type A (Wyckoff position 1) - will move in x direction
        \fill[atomTypeA] (1,1,1) circle (0.35);
        \fill[atomTypeA] (1,2,2) circle (0.35);
        
        % Atoms Type B (Wyckoff position 2) - will move in y direction
        \fill[atomTypeB] (2,0.25,1.5) circle (0.25);
        \fill[atomTypeB] (2,2.25,1.5) circle (0.25);
        
        % Atoms Type C (fixed positions)
        \fill[atomTypeC] (0.0,0.0,0.0) circle (0.2);
        \fill[atomTypeC] (0.0,0.0,3.0) circle (0.2);
        \fill[atomTypeC] (0.0,3.0,0.0) circle (0.2);
        \fill[atomTypeC] (3.0,0.0,0.0) circle (0.2);
        \fill[atomTypeC] (3.0,3.0,0.0) circle (0.2);
        \fill[atomTypeC] (0.0,3.0,3.0) circle (0.2);
        \fill[atomTypeC] (3.0,0.0,3.0) circle (0.2);
        \fill[atomTypeC] (3.0,3.0,3.0) circle (0.2);
<<<<<<< HEAD

=======
    
>>>>>>> main
        
        % Movement arrows for wyckoff positions
        \draw[-{Stealth[length=8pt]}, line width=1.5pt, atomTypeA] (1.35,1,1) -- (2.5,1,1) 
            node[midway, above, text=black, font=\large] {$x$};
        \draw[-{Stealth[length=8pt]}, line width=1.5pt, atomTypeA] (1.35,2,2) -- (2.5,2,2);
        
        \draw[-{Stealth[length=8pt]}, line width=1.5pt, atomTypeB] (2,0.25,1.85) -- (2,0.25,2.7) 
            node[midway, right, text=black, font=\large] {$y$};
        \draw[-{Stealth[length=8pt]}, line width=1.5pt, atomTypeB] (2,2.25,1.85) -- (2,2.25,2.7);
        
    \end{scope}
    \node[font=\bfseries, align=center] at (1.0,-1.8) {3D Crystal Structure};
    %\node[align=center] at (1.0,-2.4) {Atoms at same Wyckoff position\\move along specific directions};

    % Part 2: Grid of structures showing movement
    \begin{scope}[shift={(5,0)}, scale=0.75]
        \foreach \x in {0,1,2} {
            \foreach \y in {0,1,2} {
                \begin{scope}[shift={(\x*2, \y*2)}, scale=0.4]
                    \draw[black, thin] (0,0,0) -- (3,0,0) -- (3,3,0) -- (0,3,0) -- cycle;
                    \draw[black, thin] (0,0,0) -- (0,0,3) -- (3,0,3) -- (3,0,0);
                    \draw[black, thin] (0,0,3) -- (0,3,3) -- (3,3,3) -- (3,0,3);
                    \draw[black, thin] (0,3,0) -- (0,3,3);
                    \draw[black, thin] (3,3,0) -- (3,3,3);
                    
                    % Type A atoms move in x as grid x increases
                    \fill[atomTypeA] ({1+\x*0.6},1,1) circle (0.35);
                    \fill[atomTypeA] ({1+\x*0.6},2,2) circle (0.35);
                    
                    % Type B atoms move in z as grid z increases
                    \fill[atomTypeB] (2,1,{1.5+\y*0.4}) circle (0.25);
                    \fill[atomTypeB] (2,2,{1.5+\y*0.4}) circle (0.25);
                    
                    % Fixed atoms
                    \fill[atomTypeC] (0.0,0.0,0.0) circle (0.2);
                    \fill[atomTypeC] (0.0,0.0,3.0) circle (0.2);
                    \fill[atomTypeC] (0.0,3.0,0.0) circle (0.2);
                    \fill[atomTypeC] (3.0,0.0,0.0) circle (0.2);
                    \fill[atomTypeC] (3.0,3.0,0.0) circle (0.2);
                    \fill[atomTypeC] (0.0,3.0,3.0) circle (0.2);
                    \fill[atomTypeC] (3.0,0.0,3.0) circle (0.2);
                    \fill[atomTypeC] (3.0,3.0,3.0) circle (0.2);
    
                \end{scope}
                
                % Add indices below each cell
                %\node[font=\tiny] at (\x*2, \y*2-0.7) {$x={\x*0.6}, y={\y*0.4}$};
            }
        }
                
        % Add coordinate arrows for the grid
        \draw[-{Stealth[length=10pt]}, thick] (0,-1) -- (4,-1) node[right, font=\small] {$x$ position};
        \draw[-{Stealth[length=10pt]}, thick] (-1,0) -- (-1,6) node[left, font=\small] {$y$ position};
    \end{scope}

\node[font=\bfseries, align=center, scale=1] at (7.0,-1.8) {Grid of Configurations};
%\node[align=center, font=\small, scale=1] at (7.0,-2.4) {Mapping a meshgrid of degree-of-freedom \\ values to corresponding crystal structures.};

% Part 3: Energy calculation and 2D PES creation
\begin{scope}[shift={(11.5,0)}]
    \begin{axis}[
        width=6cm,
        height=6cm,
        view={0}{90},
        colormap={custom}{color(0)=(potentialCold) color(0.4)=(blue!50!black) color(0.7)=(red!50!orange) color(1)=(potentialHot)},
        colorbar right,
        colorbar style={
            title=Energy (eV),
            title style={font=\footnotesize},
            width=0.5cm
        },
        xlabel={DOF along ($x$)},
        ylabel={DOF along ($y$)},
        x tick label style={font=\footnotesize},
        y tick label style={font=\footnotesize},
        title style={font=\bfseries, align=center},
        point meta min=-2,
        point meta max=1,
        xtick={0,0.5,1.0},
        ytick={0,0.5,1.0},
    ]
    
    \addplot3[
        surf,
        domain=0:1,
        domain y=0:1,
        samples=40,
        samples y=40,
    ] {
        -0.7*exp(-((x-0.2)^2+(y-0.7)^2)/0.04) + 
        -0.9*exp(-((x-0.7)^2+(y-0.2)^2)/0.05) + 
        -0.6*exp(-((x-0.8)^2+(y-0.6)^2)/0.04) + 
        -0.4*exp(-((x-0.4)^2+(y-0.4)^2)/0.06) +
        0.5*exp(-((x-0.1)^2+(y-0.3)^2)/0.03) +
        0.8*exp(-((x-0.6)^2+(y-0.8)^2)/0.07)
    };
    
    \addplot3[only marks, mark=*, white, mark size=2pt] coordinates {
        (0.3,0.2,0.8) (0.5,0.2,0.8) (0.7,0.2,0.8)
        (0.3,0.5,0.8) (0.5,0.5,0.8) (0.7,0.5,0.8)
        (0.3,0.8,0.8) (0.5,0.8,0.8) (0.7,0.8,0.8)
    };
    
    \end{axis}
\end{scope}

\node[font=\bfseries, align=center] at (14.0,-1.8) {2D Potential Energy Surface};
%\node[align=center, font=\small, text width=6cm] at (14.0,-2.4) {Each point corresponds to a\\specific configuration in the grid};

\draw[-{Stealth[length=10pt]}, line width=1pt, draw=gray!50] (2,1) to[bend left=15] (4,1);
\draw[-{Stealth[length=10pt]}, line width=1pt, draw=gray!50] (9,1) to[bend left=10] (10.0,1);

\end{tikzpicture}

\end{center}
\vspace{+0.1in}
\caption{\textbf{Illustration of the workflow}. A toy crystal structure with three atom types is depicted: blue atoms remain fixed by symmetry,  while green and red atoms (each with multiplicity 2) can vary within $[0,1]$.
In the workflow, the user can selectively freeze or free these Wyckoff degrees of freedom.
Here, the workflow produces a 2D meshgrid \(\{(x_i, y_j) \mid x_i, y_j \in [0,1]\}\) that enumerates possible configurations in some discrete step size.
Each point on the meshgrid corresponds to a distinct crystal structure, and we calculate the energy of that structure using any interatomic potential that provides an ASE calculator interface.
The resulting 2D energy slice shows how displacements of these atoms affect the total energy of the system.
}
\label{fig:graph_repr}
\end{figure*}

%% file: references.bib
@incollection{walesEnergyLandscapes2001,
  title = {Energy {{Landscapes}}},
  booktitle = {Atomic Clusters and Nanoparticles. {{Agregats}} Atomiques et Nanoparticules: {{Les Houches Session LXXIII}} 2--28 {{July}} 2000},
  author = {Wales, D. J.},
  editor = {Guet, C. and Hobza, P. and Speigelman, F. and David, F.},
  year = {2001},
  pages = {437--507},
  publisher = {Springer},
  address = {Berlin, Heidelberg},
  url = {https://doi.org/10.1007/3-540-45621-X_10},
  urldate = {2025-04-23},
  isbn = {978-3-540-45621-6},
  langid = {english}
}

@article{Heinen_2020,
  title = {Machine Learning the Computational Cost of Quantum Chemistry},
  author = {Heinen, Stefan and Schwilk, Max and {von Rudorff}, Guido Falk and {von Lilienfeld}, O Anatole},
  year = {2020},
  month = mar,
  journal = {Machine Learning: Science and Technology},
  volume = {1},
  number = {2},
  pages = {025002},
  publisher = {IOP Publishing},
  url = {https://dx.doi.org/10.1088/2632-2153/ab6ac4}
}

@article{yuSystematicAssessmentVarious2024a,
  title = {Systematic Assessment of Various Universal Machine-Learning Interatomic Potentials},
  author = {Yu, Haochen and Giantomassi, Matteo and Materzanini, Giuliana and Wang, Junjie and Rignanese, Gian-Marco},
  year = {2024},
  journal = {Materials Genome Engineering Advances},
  volume = {2},
  number = {3},
  pages = {e58},
  issn = {2940-9497},
  url = {https://onlinelibrary.wiley.com/doi/abs/10.1002/mgea.58},
  urldate = {2025-04-23},
  copyright = {{\copyright} 2024 The Author(s). Materials Genome Engineering Advances published by Wiley-VCH GmbH on behalf of University of Science and Technology Beijing.},
  langid = {english}
}

@article{zuoPerformanceCostAssessment2020,
  title = {Performance and {{Cost Assessment}} of {{Machine Learning Interatomic Potentials}}},
  author = {Zuo, Yunxing and Chen, Chi and Li, Xiangguo and Deng, Zhi and Chen, Yiming and Behler, J{\"o}rg and Cs{\'a}nyi, G{\'a}bor and Shapeev, Alexander V. and Thompson, Aidan P. and Wood, Mitchell A. and Ong, Shyue Ping},
  year = {2020},
  month = jan,
  journal = {The Journal of Physical Chemistry A},
  volume = {124},
  number = {4},
  pages = {731--745},
  publisher = {American Chemical Society},
  issn = {1089-5639},
  url = {https://doi.org/10.1021/acs.jpca.9b08723}
}

@article{batznerE3equivariantGraphNeural2022,
  title = {E(3)-Equivariant Graph Neural Networks for Data-Efficient and Accurate Interatomic Potentials},
  author = {Batzner, Simon and Musaelian, Albert and Sun, Lixin and Geiger, Mario and Mailoa, Jonathan P. and Kornbluth, Mordechai and Molinari, Nicola and Smidt, Tess E. and Kozinsky, Boris},
  year = {2022},
  journal = {Nature Communications},
  volume = {13},
  number = {1},
  pages = {2453},
  issn = {2041-1723},
  url = {https://doi.org/10.1038/s41467-022-29939-5}
}

@article{Wen2020,
  title = {Uncertainty quantification in molecular simulations with dropout neural network potentials},
  volume = {6},
  ISSN = {2057-3960},
  url = {http://dx.doi.org/10.1038/s41524-020-00390-8},
  DOI = {10.1038/s41524-020-00390-8},
  number = {1},
  journal = {npj Computational Materials},
  publisher = {Springer Science and Business Media LLC},
  author = {Wen,  Mingjian and Tadmor,  Ellad B.},
  year = {2020},
  month = aug 
}

@article{Tran2017,
  title = {An Efficient First-Principles Saddle Point Searching Method Based on Distributed Kriging Metamodels},
  volume = {4},
  ISSN = {2332-9025},
  url = {http://dx.doi.org/10.1115/1.4037459},
  DOI = {10.1115/1.4037459},
  number = {1},
  journal = {ASCE-ASME Journal of Risk and Uncertainty in Engineering Systems,  Part B: Mechanical Engineering},
  publisher = {ASME International},
  author = {Tran,  Anh and He,  Lijuan and Wang,  Yan},
  year = {2017},
  month = sep 
}

@misc{deng2024overcomingsystematicsofteninguniversal,
      title={Overcoming systematic softening in universal machine learning interatomic potentials by fine-tuning}, 
      author={Bowen Deng and Yunyeong Choi and Peichen Zhong and Janosh Riebesell and Shashwat Anand and Zhuohan Li and KyuJung Jun and Kristin A. Persson and Gerbrand Ceder},
      year={2024},
      eprint={2405.07105},
      archivePrefix={arXiv},
      primaryClass={cond-mat.mtrl-sci},
      url={https://arxiv.org/abs/2405.07105}, 
}

@article{chgnet,
  title = {CHGNet as a pretrained universal neural network potential for charge-informed atomistic modelling},
  volume = {5},
  ISSN = {2522-5839},
  url = {http://dx.doi.org/10.1038/s42256-023-00716-3},
  DOI = {10.1038/s42256-023-00716-3},
  number = {9},
  journal = {Nature Machine Intelligence},
  publisher = {Springer Science and Business Media LLC},
  author = {Deng,  Bowen and Zhong,  Peichen and Jun,  KyuJung and Riebesell,  Janosh and Han,  Kevin and Bartel,  Christopher J. and Ceder,  Gerbrand},
  year = {2023},
  month = sep,
  pages = {1031–1041}
}

@article{jainCommentaryMaterialsProject2013,
  title = {Commentary: {{The Materials Project}}: {{A}} Materials Genome Approach to Accelerating Materials Innovation},
  shorttitle = {Commentary},
  author = {Jain, Anubhav and Ong, Shyue Ping and Hautier, Geoffroy and Chen, Wei and Richards, William Davidson and Dacek, Stephen and Cholia, Shreyas and Gunter, Dan and Skinner, David and Ceder, Gerbrand and Persson, Kristin A.},
  year = 2013,
  month = jul,
  journal = {APL Materials},
  volume = {1},
  number = {1},
  pages = {011002},
  publisher = {American Institute of Physics},
  url = {https://aip.scitation.org/doi/10.1063/1.4812323},
  urldate = {2022-12-05}
}

@misc{yangMatterSimDeepLearning2024,
  title = {{{MatterSim}}: {{A Deep Learning Atomistic Model Across Elements}}, {{Temperatures}} and {{Pressures}}},
  shorttitle = {{{MatterSim}}},
  author = {Yang, Han and Hu, Chenxi and Zhou, Yichi and Liu, Xixian and Shi, Yu and Li, Jielan and Li, Guanzhi and Chen, Zekun and Chen, Shuizhou and Zeni, Claudio and Horton, Matthew and Pinsler, Robert and Fowler, Andrew and Z{\"u}gner, Daniel and Xie, Tian and Smith, Jake and Sun, Lixin and Wang, Qian and Kong, Lingyu and Liu, Chang and Hao, Hongxia and Lu, Ziheng},
  year = 2024,
  month = may,
  number = {arXiv:2405.04967},
  eprint = {2405.04967},
  primaryclass = {cond-mat},
  publisher = {arXiv},
  url = {http://arxiv.org/abs/2405.04967},
  urldate = {2025-10-20},
  archiveprefix = {arXiv}
}

@article{merchantScalingDeepLearning2023,
  title = {Scaling Deep Learning for Materials Discovery},
  author = {Merchant, Amil and Batzner, Simon and Schoenholz, Samuel S. and Aykol, Muratahan and Cheon, Gowoon and Cubuk, Ekin Dogus},
  year = 2023,
  month = dec,
  journal = {Nature},
  volume = {624},
  number = {7990},
  pages = {80--85},
  publisher = {Nature Publishing Group},
  issn = {1476-4687},
  url = {https://www.nature.com/articles/s41586-023-06735-9},
  urldate = {2024-02-12},
  copyright = {2023 The Author(s)},
  langid = {english}
}

@article{erhardModellingAtomicNanoscale2024,
  title = {Modelling Atomic and Nanoscale Structure in the Silicon--Oxygen System through Active Machine Learning},
  author = {Erhard, Linus C. and Rohrer, Jochen and Albe, Karsten and Deringer, Volker L.},
  year = 2024,
  month = mar,
  journal = {Nature Communications},
  volume = {15},
  number = {1},
  pages = {1927},
  publisher = {Nature Publishing Group},
  issn = {2041-1723},
  url = {https://www.nature.com/articles/s41467-024-45840-9},
  urldate = {2025-10-20},
  copyright = {2024 The Author(s)},
  langid = {english}
}

@misc{mace,
  doi = {10.48550/ARXIV.2401.00096},
  url = {https://arxiv.org/abs/2401.00096},
  author = {Batatia,  Ilyes and Benner,  Philipp and Chiang,  Yuan and Elena,  Alin M. and Kovács,  Dávid P. and Riebesell,  Janosh and Advincula,  Xavier R. and Asta,  Mark and Avaylon,  Matthew and Baldwin,  William J. and Berger,  Fabian and Bernstein,  Noam and Bhowmik,  Arghya and Blau,  Samuel M. and Cărare,  Vlad and Darby,  James P. and De,  Sandip and Della Pia,  Flaviano and Deringer,  Volker L. and Elijošius,  Rokas and El-Machachi,  Zakariya and Falcioni,  Fabio and Fako,  Edvin and Ferrari,  Andrea C. and Genreith-Schriever,  Annalena and George,  Janine and Goodall,  Rhys E. A. and Grey,  Clare P. and Grigorev,  Petr and Han,  Shuang and Handley,  Will and Heenen,  Hendrik H. and Hermansson,  Kersti and Holm,  Christian and Jaafar,  Jad and Hofmann,  Stephan and Jakob,  Konstantin S. and Jung,  Hyunwook and Kapil,  Venkat and Kaplan,  Aaron D. and Karimitari,  Nima and Kermode,  James R. and Kroupa,  Namu and Kullgren,  Jolla and Kuner,  Matthew C. and Kuryla,  Domantas and Liepuoniute,  Guoda and Margraf,  Johannes T. and Magdău,  Ioan-Bogdan and Michaelides,  Angelos and Moore,  J. Harry and Naik,  Aakash A. and Niblett,  Samuel P. and Norwood,  Sam Walton and O'Neill,  Niamh and Ortner,  Christoph and Persson,  Kristin A. and Reuter,  Karsten and Rosen,  Andrew S. and Schaaf,  Lars L. and Schran,  Christoph and Shi,  Benjamin X. and Sivonxay,  Eric and Stenczel,  Tamás K. and Svahn,  Viktor and Sutton,  Christopher and Swinburne,  Thomas D. and Tilly,  Jules and van der Oord,  Cas and Varga-Umbrich,  Eszter and Vegge,  Tejs and Vondrák,  Martin and Wang,  Yangshuai and Witt,  William C. and Zills,  Fabian and Csányi,  Gábor},
  title = {A foundation model for atomistic materials chemistry},
  publisher = {arXiv},
  year = {2024},
  copyright = {Creative Commons Attribution Non Commercial No Derivatives 4.0 International}
}

@article{liuDiscrepanciesErrorEvaluation2023,
  title = {Discrepancies and Error Evaluation Metrics for Machine Learning Interatomic Potentials},
  author = {Liu, Yunsheng and He, Xingfeng and Mo, Yifei},
  year = {2023},
  month = sep,
  journal = {npj Computational Materials},
  volume = {9},
  number = {1},
  pages = {1--13},
  publisher = {Nature Publishing Group},
  issn = {2057-3960},
  url = {https://www.nature.com/articles/s41524-023-01123-3},
  urldate = {2025-04-25},
  copyright = {2023 The Author(s)},
  langid = {english}
}

@misc{neumannOrbFastScalable2024,
  title = {Orb: {{A Fast}}, {{Scalable Neural Network Potential}}},
  shorttitle = {Orb},
  author = {Neumann, Mark and Gin, James and Rhodes, Benjamin and Bennett, Steven and Li, Zhiyi and Choubisa, Hitarth and Hussey, Arthur and Godwin, Jonathan},
  year = {2024},
  month = oct,
  number = {arXiv:2410.22570},
  eprint = {2410.22570},
  primaryclass = {cond-mat},
  publisher = {arXiv},
  url = {http://arxiv.org/abs/2410.22570},
  urldate = {2025-04-24},
  archiveprefix = {arXiv}
}

@article{sevenet,
	title = {Scalable Parallel Algorithm for Graph Neural Network Interatomic Potentials in Molecular Dynamics Simulations},
	volume = {20},
	doi = {10.1021/acs.jctc.4c00190},
	number = {11},
	journal = {J. Chem. Theory Comput.},
	author = {Park, Yutack and Kim, Jaesun and Hwang, Seungwoo and Han, Seungwu},
	year = {2024},
	pages = {4857--4868},
}

@book{wyckoff1922analytical,
  title={The Analytical Expression of the Results of the Theory of Space-groups},
  author={Wyckoff, Ralph Walter Graystone},
  number={318},
  year={1922},
  publisher={Carnegie institution of Washington}
}

@article{goodallRapidDiscoveryStable2022b,
	title = {Rapid discovery of stable materials by coordinate-free coarse graining},
	volume = {8},
	url = {https://www.science.org/doi/full/10.1126/sciadv.abn4117},
	doi = {10.1126/sciadv.abn4117},
	number = {30},
	urldate = {2022-08-15},
	journal = {Science Advances},
	author = {Goodall, Rhys E. A. and Parackal, Abhijith S. and Faber, Felix A. and Armiento, Rickard and Lee, Alpha A.},
	month = jul,
	year = {2022},
	note = {Publisher: American Association for the Advancement of Science},
	pages = {eabn4117},
}

@article{Wang2018,
  title = {Expedite random structure searching using objects from Wyckoff positions},
  volume = {148},
  ISSN = {1089-7690},
  url = {http://dx.doi.org/10.1063/1.5006104},
  DOI = {10.1063/1.5006104},
  number = {5},
  journal = {The Journal of Chemical Physics},
  publisher = {AIP Publishing},
  author = {Wang,  Shu-Wei and Hsing,  Cheng-Rong and Wei,  Ching-Ming},
  year = {2018},
  month = feb 
}

@article{xrd_paper,
  title = {Identifying crystal structures beyond known prototypes from x-ray powder diffraction spectra},
  author = {Parackal, Abhijith S. and Goodall, Rhys E. A. and Faber, Felix A. and Armiento, Rickard},
  journal = {Phys. Rev. Mater.},
  volume = {8},
  issue = {10},
  pages = {103801},
  numpages = {14},
  year = {2024},
  month = {Oct},
  publisher = {American Physical Society},
  doi = {10.1103/PhysRevMaterials.8.103801},
  url = {https://link.aps.org/doi/10.1103/PhysRevMaterials.8.103801}
}

@article{lenz2019parametrically,
  title={Parametrically constrained geometry relaxations for high-throughput materials science},
  author={Lenz, Maja-Olivia and Purcell, Thomas AR and Hicks, David and Curtarolo, Stefano and Scheffler, Matthias and Carbogno, Christian},
  journal={npj Computational Materials},
  volume={5},
  number={1},
  pages={123},
  year={2019},
  publisher={Nature Publishing Group UK London}
}

@article{Reinaudi2000,
  title = {Simulated annealing prediction of the crystal structure of ternary inorganic compounds using symmetry restrictions †},
  ISSN = {1364-5447},
  url = {http://dx.doi.org/10.1039/B003447I},
  DOI = {10.1039/b003447i},
  number = {23},
  journal = {Journal of the Chemical Society,  Dalton Transactions},
  publisher = {Royal Society of Chemistry (RSC)},
  author = {Reinaudi,  Luis and Leiva,  Ezequiel P. M. and Carbonio,  Raúl E.},
  year = {2000},
  pages = {4258–4262}
}

@article{AFLOWLibraryCrystallographic,
  title = {The {{AFLOW}} Library of Crystallographic Prototypes: {{Part}} 1},
  author = {Mehl, Michael J. and Hicks, David and Toher, Cormac and Levy, Ohad and Hanson, Robert M. and Hart, Gus and Curtarolo, Stefano},
  year = {2017},
  journal = {Comput. Mater. Sci.},
  volume = {136},
  pages = {S1--S828},
  issn = {0927-0256},
  doi = {10.1016/j.commatsci.2017.01.017}
}

@article{donaldson2024genetic,
  title={A Genetic Algorithm For Convex Hull Optimisation},
  author={Donaldson, Scott and Lawrence, Robert A and Probert, Matt IJ},
  journal={arXiv preprint arXiv:2404.14354},
  year={2024}
}

@article{Eyert2023,
  title = {Machine-learned interatomic potentials: Recent developments and prospective applications},
  volume = {38},
  ISSN = {2044-5326},
  url = {http://dx.doi.org/10.1557/s43578-023-01239-8},
  DOI = {10.1557/s43578-023-01239-8},
  number = {24},
  journal = {Journal of Materials Research},
  publisher = {Springer Science and Business Media LLC},
  author = {Eyert,  Volker and Wormald,  Jonathan and Curtin,  William A. and Wimmer,  Erich},
  year = {2023},
  month = dec,
  pages = {5079–5094}
}

@misc{riebesellMatbenchDiscoveryFramework2024,
  title = {Matbench {{Discovery}} -- {{A}} Framework to Evaluate Machine Learning Crystal Stability Predictions},
  author = {Riebesell, Janosh and Goodall, Rhys E. A. and Benner, Philipp and Chiang, Yuan and Deng, Bowen and Ceder, Gerbrand and Asta, Mark and Lee, Alpha A. and Jain, Anubhav and Persson, Kristin A.},
  year = {2024},
  month = dec,
  number = {arXiv:2308.14920},
  eprint = {2308.14920},
  primaryclass = {cond-mat},
  publisher = {arXiv},
  url = {http://arxiv.org/abs/2308.14920},
  urldate = {2025-04-23},
  archiveprefix = {arXiv}
}

@article{ase-paper,
  author={Ask Hjorth Larsen and Jens Jørgen Mortensen and Jakob Blomqvist and Ivano E Castelli and Rune Christensen and Marcin
Dułak and Jesper Friis and Michael N Groves and Bjørk Hammer and Cory Hargus and Eric D Hermes and Paul C Jennings and Peter
Bjerre Jensen and James Kermode and John R Kitchin and Esben Leonhard Kolsbjerg and Joseph Kubal and Kristen
Kaasbjerg and Steen Lysgaard and Jón Bergmann Maronsson and Tristan Maxson and Thomas Olsen and Lars Pastewka and Andrew
Peterson and Carsten Rostgaard and Jakob Schiøtz and Ole Schütt and Mikkel Strange and Kristian S Thygesen and Tejs
Vegge and Lasse Vilhelmsen and Michael Walter and Zhenhua Zeng and Karsten W Jacobsen},
  title={The atomic simulation environment—a Python library for working with atoms},
  journal={Journal of Physics: Condensed Matter},
  volume={29},
  number={27},
  pages={273002},
  url={http://stacks.iop.org/0953-8984/29/i=27/a=273002},
  year={2017}
}

@article{spglibv2,
  author = "Shinohara, Kohei and Togo, Atsushi and Tanaka, Isao",
  title = "{Algorithms for magnetic symmetry operation search and identification of magnetic space group from magnetic crystal structure}",
  journal = "Acta Cryst. A",
  year = "2023",
  volume = "79",
  number = "5",
  pages = "390--398",
  month = "Sep",
  doi = {10.1107/S2053273323005016},
  url = {https://doi.org/10.1107/S2053273323005016},
}

@software{jax2018github,
  author = {James Bradbury and Roy Frostig and Peter Hawkins and Matthew James Johnson and Chris Leary and Dougal Maclaurin and George Necula and Adam Paszke and Jake Vander{P}las and Skye Wanderman-{M}ilne and Qiao Zhang},
  title = {{JAX}: composable transformations of {P}ython+{N}um{P}y programs},
  url = {http://github.com/jax-ml/jax},
  version = {0.3.13},
  year = {2018},
}

@article{PhysRevX.11.041026,
  title = {Visualizing Energy Landscapes through Manifold Learning},
  author = {Shires, Benjamin W. B. and Pickard, Chris J.},
  journal = {Phys. Rev. X},
  volume = {11},
  issue = {4},
  pages = {041026},
  numpages = {23},
  year = {2021},
  month = {Nov},
  publisher = {American Physical Society},
  doi = {10.1103/PhysRevX.11.041026},
  url = {https://link.aps.org/doi/10.1103/PhysRevX.11.041026}
}

@article{smeetonVisualizingEnergyLandscapes2014a,
	title = {Visualizing energy landscapes with metric disconnectivity graphs},
	volume = {35},
	issn = {1096-987X},
	url = {https://onlinelibrary.wiley.com/doi/abs/10.1002/jcc.23643},
	doi = {10.1002/jcc.23643},
	language = {en},
	number = {20},
	urldate = {2025-04-23},
	journal = {Journal of Computational Chemistry},
	author = {Smeeton, Lewis C. and Oakley, Mark T. and Johnston, Roy L.},
	year = {2014},
	note = {\_eprint: https://onlinelibrary.wiley.com/doi/pdf/10.1002/jcc.23643},
	pages = {1481--1490},
}

@article{walesDecodingEnergyLandscape2012,
  title = {Decoding the Energy Landscape: Extracting Structure, Dynamics and Thermodynamics},
  shorttitle = {Decoding the Energy Landscape},
  author = {Wales, David J.},
  year = {2012},
  month = jun,
  journal = {Philosophical Transactions of the Royal Society A: Mathematical, Physical and Engineering Sciences},
  volume = {370},
  number = {1969},
  pages = {2877--2899},
  publisher = {Royal Society},
  url = {https://royalsocietypublishing.org/doi/10.1098/rsta.2011.0208},
  urldate = {2025-04-23}
}

@inproceedings{
bihani2024lowdimensional,
title={Low-Dimensional Projections for Visualizing Energy Landscapes of Atomic Systems},
author={Vaibhav Bihani and Srikanth Sastry and Sayan Ranu and N M Anoop Krishnan},
booktitle={AI for Accelerated Materials Design - Vienna 2024},
year={2024},
url={https://openreview.net/forum?id=x3ryxZgHgu}
}

@article{Müser31122023,
author = {Martin H. Müser and Sergey V. Sukhomlinov and Lars Pastewka},
title = {Interatomic potentials: achievements and challenges},
journal = {Advances in Physics: X},
volume = {8},
number = {1},
pages = {2093129},
year = {2023},
publisher = {Taylor \& Francis},
doi = {10.1080/23746149.2022.2093129},
URL = {https://doi.org/10.1080/23746149.2022.2093129},
eprint = {https://doi.org/10.1080/23746149.2022.2093129}
}

@article{kaurDataefficientFinetuningFoundational2025,
  title = {Data-Efficient Fine-Tuning of Foundational Models for First-Principles Quality Sublimation Enthalpies},
  author = {Kaur, Harveen and Pia, Flaviano Della and Batatia, Ilyes and Advincula, Xavier R. and Shi, Benjamin X. and Lan, Jinggang and Cs{\'a}nyi, G{\'a}bor and Michaelides, Angelos and Kapil, Venkat},
  year = 2025,
  month = jan,
  journal = {Faraday Discussions},
  volume = {256},
  number = {0},
  pages = {120--138},
  publisher = {The Royal Society of Chemistry},
  issn = {1364-5498},
  url = {https://pubs.rsc.org/en/content/articlelanding/2025/fd/d4fd00107a},
  urldate = {2025-10-20},
  langid = {english}
}

@misc{kumarFineTuningCanDistort2022,
  title = {Fine-{{Tuning}} Can {{Distort Pretrained Features}} and {{Underperform Out-of-Distribution}}},
  author = {Kumar, Ananya and Raghunathan, Aditi and Jones, Robbie and Ma, Tengyu and Liang, Percy},
  year = 2022,
  month = feb,
  number = {arXiv:2202.10054},
  eprint = {2202.10054},
  primaryclass = {cs},
  publisher = {arXiv},
  url = {http://arxiv.org/abs/2202.10054},
  urldate = {2025-10-20},
  archiveprefix = {arXiv}
}

@article{radovaFinetuningFoundationModels2025,
  title = {Fine-Tuning Foundation Models of Materials Interatomic Potentials with Frozen Transfer Learning},
  author = {Radova, Mariia and Stark, Wojciech G. and Allen, Connor S. and Maurer, Reinhard J. and Bart{\'o}k, Albert P.},
  year = 2025,
  month = jul,
  journal = {npj Computational Materials},
  volume = {11},
  number = {1},
  pages = {237},
  publisher = {Nature Publishing Group},
  issn = {2057-3960},
  url = {https://www.nature.com/articles/s41524-025-01727-x},
  urldate = {2025-10-20},
  copyright = {2025 The Author(s)},
  langid = {english}
}

@article{schmidtMachineLearningAssistedDeterminationGlobal2023,
  title = {Machine-{{Learning-Assisted Determination}} of the {{Global Zero-Temperature Phase Diagram}} of {{Materials}}},
  author = {Schmidt, Jonathan and Hoffmann, Noah and Wang, Hai-Chen and Borlido, Pedro and Carri{\c c}o, Pedro J. M. A. and Cerqueira, Tiago F. T. and Botti, Silvana and Marques, Miguel A. L.},
  year = {2023},
  journal = {Advanced Materials},
  volume = {35},
  number = {22},
  pages = {2210788},
  issn = {1521-4095},
  url = {https://onlinelibrary.wiley.com/doi/abs/10.1002/adma.202210788},
  urldate = {2025-04-24},
  copyright = {{\copyright} 2023 The Authors. Advanced Materials published by Wiley-VCH GmbH},
  langid = {english}
}

@misc{huangCrossfunctionalTransferabilityUniversal2025,
  title = {Cross-Functional Transferability in Universal Machine Learning Interatomic Potentials},
  author = {Huang, Xu and Deng, Bowen and Zhong, Peichen and Kaplan, Aaron D. and Persson, Kristin A. and Ceder, Gerbrand},
  year = {2025},
  month = apr,
  number = {arXiv:2504.05565},
  eprint = {2504.05565},
  primaryclass = {cond-mat},
  publisher = {arXiv},
  url = {http://arxiv.org/abs/2504.05565},
  urldate = {2025-04-24},
  archiveprefix = {arXiv}
}

@misc{kaplanFoundationalPotentialEnergy2025,
  title = {A {{Foundational Potential Energy Surface Dataset}} for {{Materials}}},
  author = {Kaplan, Aaron D. and Liu, Runze and Qi, Ji and Ko, Tsz Wai and Deng, Bowen and Riebesell, Janosh and Ceder, Gerbrand and Persson, Kristin A. and Ong, Shyue Ping},
  year = {2025},
  month = mar,
  number = {arXiv:2503.04070},
  eprint = {2503.04070},
  primaryclass = {cond-mat},
  publisher = {arXiv},
  url = {http://arxiv.org/abs/2503.04070},
  urldate = {2025-04-24},
  archiveprefix = {arXiv}
}

@misc{OMat24,
  title = {Open {{Materials}} 2024 ({{OMat24}}) {{Inorganic Materials Dataset}} and {{Models}}},
  author = {{Barroso-Luque}, Luis and Shuaibi, Muhammed and Fu, Xiang and Wood, Brandon M. and Dzamba, Misko and Gao, Meng and Rizvi, Ammar and Zitnick, C. Lawrence and Ulissi, Zachary W.},
  year = {2024},
  month = oct,
  number = {arXiv:2410.12771},
  eprint = {2410.12771},
  primaryclass = {cond-mat},
  publisher = {arXiv},
  url = {http://arxiv.org/abs/2410.12771},
  urldate = {2025-04-24},
  archiveprefix = {arXiv}
}

@misc{mace_mp_0,
	author = {},
	title = {{MACE-MP-0}},
	howpublished = {\url{https://github.com/ACEsuit/mace-foundations/tree/mace_mp_0}},
	year = {2024},
}

@misc{mace_mp_0b2,
	author = {},
	title = {{MACE-MP-0b2}},
	howpublished = {\url{https://github.com/ACEsuit/mace-foundations/tree/mace_mp_0b2}},
	year = {2024},
}

@misc{mace_mp_0b3,
	author = {},
	title = {{MACE-MP-0b3}},
	howpublished = {\url{https://github.com/ACEsuit/mace-foundations/tree/mace_mp_0b3}},
	year = {2024},

}

@misc{mace_mpa_0,
	author = {},
	title = {{MACE-MPA-0}},
	howpublished = {\url{https://github.com/ACEsuit/mace-foundations/tree/mace_mpa_0}},
	year = {2024},

}

@misc{mace_omat_0,
	author = {},
	title = {{MACE-OMAT-0}},
	howpublished = {\url{https://github.com/ACEsuit/mace-foundations/tree/mace_omat_0}},
	year = {2025},

}

@misc{mace_matpes_0,
	author = {},
	title = {{MACE-MatPES-0}},
	howpublished = {\url{https://github.com/ACEsuit/mace-foundations/tree/mace_matpes_0}},
	year = {2025},
}

@misc{sevenet_pretrained_0,
	author = {Park, Yutack and Kim, Jaesun and Hwang, Seungwoo and Han, Seungwu},
	year = {2024},
    title = {{S}even{N}et},
	howpublished = {\url{https://github.com/MDIL-SNU/SevenNet/tree/v0.10.1/sevenn/pretrained_potentials/SevenNet_0__11July2024}},
	note = {[Accessed 07-05-2025]},
}

@misc{orb_v2,
    author={Mark Neumann and James Gin and Benjamin Rhodes and Steven Bennett and Zhiyi Li and Hitarth Choubisa and Arthur Hussey and Jonathan Godwin},
    title = {ORB v2 weights},
	howpublished = {\url{https://orbitalmaterials-public-models.s3.us-west-1.amazonaws.com/forcefields/orb-v2-20241011.ckpt}},
	note = {[Accessed 07-05-2025]},
    year={2024},
}

@misc{chgnet_030,
    author={Deng, Bowen and Zhong, Peichen and Jun, KyuJung and Riebesell, Janosh and Han, Kevin and Bartel, Christopher J. and Ceder, Gerbrand},
	title = {chgnet},
	howpublished = {\url{https://github.com/CederGroupHub/chgnet/tree/main/chgnet/pretrained/0.3.0}},
	note = {[Accessed 07-05-2025]},
    year={2023}
}

@article{Hohenberg1964,
  title = {Inhomogeneous Electron Gas},
  volume = {136},
  ISSN = {0031-899X},
  url = {http://dx.doi.org/10.1103/PhysRev.136.B864},
  DOI = {10.1103/physrev.136.b864},
  number = {3B},
  journal = {Physical Review},
  publisher = {American Physical Society (APS)},
  author = {Hohenberg,  P. and Kohn,  W.},
  year = {1964},
  month = nov,
  pages = {B864–B871}
}

@article{Kohn1965,
  title = {Self-Consistent Equations Including Exchange and Correlation Effects},
  volume = {140},
  ISSN = {0031-899X},
  url = {http://dx.doi.org/10.1103/PhysRev.140.A1133},
  DOI = {10.1103/physrev.140.a1133},
  number = {4A},
  journal = {Physical Review},
  publisher = {American Physical Society (APS)},
  author = {Kohn,  W. and Sham,  L. J.},
  year = {1965},
  month = nov,
  pages = {A1133–A1138}
}

@misc{parackalScreening39Billion2026,
  title = {Screening 39 Billion Protostructures for Materials Discovery},
  author = {Parackal, Abhijith S. and Trybel, Florian and Faber, Felix Andreas and Armiento, Rickard},
  year = 2026,
  month = jan,
  number = {arXiv:2601.21393},
  eprint = {2601.21393},
  primaryclass = {cond-mat},
  publisher = {arXiv},
  url = {http://arxiv.org/abs/2601.21393},
  urldate = {2026-01-30},
  archiveprefix = {arXiv}
}

@book{ITA2002,
  address = {Dordrecht, Boston, London},
  author = {IUCr},
  edition = {5. revised edition},
  publisher = {Kluwer Academic Publishers},
  series = {International Tables for Crystallography},
  title = {International Tables for Crystallography, Volume A: Space Group Symmetry},
  year = 2002
}

@article{doi:10.1126/sciadv.abi7948,
author = {Jonathan Schmidt  and Love Pettersson  and Claudio Verdozzi  and Silvana Botti  and Miguel A. L. Marques },
title = {Crystal graph attention networks for the prediction of stable materials},
journal = {Science Advances},
volume = {7},
number = {49},
pages = {eabi7948},
year = {2021},
doi = {10.1126/sciadv.abi7948},
URL = {https://www.science.org/doi/abs/10.1126/sciadv.abi7948},
eprint = {https://www.science.org/doi/pdf/10.1126/sciadv.abi7948}}

@article{schmidt2024,
  title = {Improving Machine-Learning Models in Materials Science through Large Datasets},
  author = {Schmidt, Jonathan and Cerqueira, Tiago F.T. and Romero, Aldo H. and Loew, Antoine and J{\"a}ger, Fabian and Wang, Hai-Chen and Botti, Silvana and Marques, Miguel A.L.},
  year = 2024,
  journal = {Materials Today Physics},
  volume = {48},
  pages = {101560},
  issn = {2542-5293},
  url = {https://www.sciencedirect.com/science/article/pii/S2542529324002360}
}

@article{Liang2024,
  title = {High‐Pressure Synthesis of Ultra‐Incompressible,  Hard and Superconducting Tungsten Nitrides},
  volume = {34},
  ISSN = {1616-3028},
  url = {http://dx.doi.org/10.1002/adfm.202313819},
  DOI = {10.1002/adfm.202313819},
  number = {32},
  journal = {Advanced Functional Materials},
  publisher = {Wiley},
  author = {Liang,  Akun and Osmond,  Israel and Krach,  Georg and Shi,  Lan‐Ting and Br\"{u}ning,  Lukas and Ranieri,  Umbertoluca and Spender,  James and Tasnadi,  Ferenc and Massani,  Bernhard and Stevens,  Callum R. and McWilliams,  Ryan Stewart and Bright,  Eleanor Lawrence and Giordano,  Nico and Gallego‐Parra,  Samuel and Yin,  Yuqing and Aslandukov,  Andrey and Akbar,  Fariia Iasmin and Gregoryanz,  Eugene and Huxley,  Andrew and Peña‐Alvarez,  Miriam and Si,  Jian‐Guo and Schnick,  Wolfgang and Bykov,  Maxim and Trybel,  Florian and Laniel,  Dominique},
  year = {2024},
  month = may 
}

@misc{macebodyorder,
   author       = {Batatia, Ilyes},
   title        = {{MACE} Users Tutorial Notebook},
   year         = {2023},
   howpublished = {\url{https://github.com/ilyes319/mace-tutorials/blob/main/mace-users/MACE_users.ipynb}},
   note         = {Accessed: 2026-01-16},
}
